\documentclass[a4paper,fleqn,usenatbib]{mnras}

\usepackage[T1]{fontenc}
\usepackage{ae,aecompl}

\usepackage[dvipdfmx]{graphicx}	
\usepackage{amsmath}	
\usepackage{amssymb}	
\usepackage{multicol}
\usepackage{bmpsize}
\usepackage{color}

\def\startdata{\if@table@not@headed\kill\caption{\\%
    \@tablecaption}\endhead\hline\endfoot%
  \fi%
}

\def\enddata{%
 \crcr 
 \noalign{\vskip .7ex}%
 \before@enddata 
 \endtabular 
 \setbox\pt@box\lastbox 
 \pt@width\wd\pt@box\box\pt@box 
}%

\title[The high energy irradiation of HD\,209458b]{Reconstructing the high energy irradiation of the evaporating hot Jupiter HD\,209458b}

\author[T. Louden, P. J. Wheatley \& K. Briggs]{Tom Louden$^{1}$\thanks{E-mail: t.m.louden@warwick.ac.uk}, Peter J. Wheatley$^{1}$ and Kevin Briggs\\
$^{1}$Department of Physics, University of Warwick, Coventry, CV4 7AL, UK}

\date{Accepted XXX. Received YYY; in original form ZZZ}

\pubyear{2015}

\begin{document}
\label{firstpage}
\pagerange{\pageref{firstpage}--\pageref{lastpage}}
\maketitle

\begin{abstract}
The atmosphere of the exoplanet HD\,209458b is undergoing sustained mass loss, believed to be caused by X-ray and extreme-ultraviolet (XUV) irradiation from its star. The majority of this flux is not directly observable due to interstellar absorption, but is required in order to correctly model the photo-evaporation of the planet and photo-ionisation of the outflow. We present a recovered high energy spectrum for HD\,209458 using a Differential Emission Measure (DEM) retrieval technique. We construct a model of the stellar corona and transition region for temperatures between 10$^{4.1}$ and 10$^{8}$ K which is constrained jointly by ultraviolet line strengths measured with the Cosmic Origins Spectrograph (COS) on the \emph{Hubble Space Telescope} (\emph{HST}) and X-ray flux measurements from \emph{XMM-Newton}. The total hydrogen ionising luminosity ($\lambda < 912$ \AA) is found to be 10$^{28.26}$ erg s$^{-1}$, which is similar to the value for the mean activity level of the Sun. This luminosity is incompatible with energy limited mass loss rates estimated from the same COS dataset, even the lower bound requires an uncomfortably high energetic efficiency of >40\%. However, our luminosity is compatible with early estimates of the mass loss rate of HD\,209458b based on results from the \emph{HST} Space Telescope Imaging Spectrograph (STIS). Precisely reconstructed XUV irradiation is a key input to determining mass loss rates and efficiencies for exoplanet atmospheres.
\end{abstract}




\section{Introduction}
Observations of the transit of HD\,209458b in the ultraviolet show that the planet occults a much larger fraction of the star than in the optical, with a 15\% loss of flux reported in the Ly\,$\alpha$ line with \emph{Hubble Space Telescope} (\emph{HST}) Space Telescope Imaging Spectrograph (STIS) \citep{Vidal-Madjar2003}. This implies a cloud of optically thick neutral hydrogen that extends further than the Roche lobe of the planet, indicating a significant exosphere of escaping hydrogen. The presence of such an extended atmosphere, where it would be vulnerable to radiation pressure and stellar wind, implies a mass-loss rate on the order of $10^{10}$g s$^{-1}$. Further strengthening this conclusion, the occulted Ly\,$\alpha$ line showed evidence of velocity structure, with velocities between -130 and +100 km s$^{-1}$ having the strongest signal. These observations have been repeated with lower resolution instruments \citep{Vidal-Madjar2004,Ehrenreich2008,Vidal-Madjar2008}, whose results are consistent with this conclusion, though they were unable to improve the mass-loss estimates or the velocity structure of the wind.

The exosphere has also been shown to contain heavier elements, \mbox{O\,{\sc i}} and \mbox{C\,{\sc ii}} absorption have both been detected in transit \citep{Vidal-Madjar2004}. This indicates a much more dramatic mass-loss, since heavier elements should not be present in detectable quantities this high in the exosphere through molecular diffusion alone. Rather than Jeans escape, a more plausible scenario is a hydrodynamic blow off, sweeping the heavier elements with it into space. Using the then newly installed Cosmic Origins Spectrograph (COS) instrument on the \emph{HST} \cite{Linsky2010} also detected a significant outflow of ionized carbon and silicon, with transit depths of {\raise.17ex\hbox{$\scriptstyle\sim$}}8\%. Using a spherically-symmetric evaporation model, they inferred a mass-loss rate between 8 and 40 $ \times 10^{10}$ g s$^{-1}$. This higher evaporation rate uses several independent lines and avoids Ly\,$\alpha$ which suffers from both heavy interstellar medium absorption at low velocities and strong time varying geocoronal contamination, particularly with the slitless COS instrument. Whilst their carbon transit depth was consistent with that of \citet{Vidal-Madjar2004}, their detection of silicon absorption was inconsistent at the 2$\sigma$ level. Revisiting the system in the FUV, \citet{Ballester2015} find variability in the stellar emission of \mbox{Si\,{\sc iii}} and \mbox{C\,{\sc ii}} which was not accounted for in \citet{Linsky2010}. They find no evidence for \mbox{Si\,{\sc iii}} absorption, and slightly smaller absorption of \mbox{C\,{\sc ii}}, so the mass-loss rate is most likely smaller than that reported by \cite{Linsky2010}.

There is a consensus that HD\,209458b is undergoing large-scale atmospheric evaporation, believed to be caused by X-ray and EUV (XUV) heating of the planetary atmosphere \citep{Lammer2003}. The optical flux of the star is not capable of heating the atmosphere to the {\raise.17ex\hbox{$\scriptstyle\sim$}}10,000 K temperature required to outflow from the planet, but XUV photons from the stellar corona and transition region have enough energy to ionize a hydrogen atom ($\lambda$ < 912 \AA), releasing cascades of electrons that quickly heat the gas to extremely high temperatures. Unfortunately the neutral and molecular hydrogen of the ISM is very efficient at blocking radiation at these wavelengths, so it is very difficult to observe on all but the very closest of stars. There is also a current lack of space based observatories capable of imaging in this wavelength region.

X-ray observations can characterize the hotter regions of a stellar corona, but for solar type stars over 90\% of the ionising flux is emitted in the EUV (100 < $\lambda$ < 912 \AA) which is mainly emitted from cooler regions, between 10$^{4.5}$ and 10$^{6.5}$ K. In the absence of direct observations of the EUV flux of the star, a model of the corona is necessary for the full high energy spectrum to be recovered. Fortunately, high excitation ultraviolet lines exist that are formed in the same temperature regions of the corona as the EUV flux and can be used as a proxy for ionising flux. These lines have high ionisation states, so are only minimally effected by ISM absorption. We combine measurements of these UV lines with \emph{XMM-Newton} observations to construct a model of the corona which we use to recover the entire high energy spectrum of HD\,209458. 

\citet{Linsky2014} present a technique for calculating the EUV luminosity of stars by scaling Ly\,$\alpha$ luminosity. Since for many exoplanet host stars the majority of the Ly\,$\alpha$ flux will be absorbed by the interstellar medium, it has to be reconstructed, which could introduce uncertainties to this technique. Although their relationship is based on a small sample that includes only 4 G class stars, and despite the uncertainties of Ly\,$\alpha$ flux reconstruction, this method can be a useful tool for estimating the EUV fluxes of planet-hosting stars on a population level, but the {\raise.17ex\hbox{$\scriptstyle\sim$}}30\% dispersion of stars around their fit means DEM recovery is more appropriate for detailed analysis of individual evaporating systems such as HD\,209458b. An accurate determination of the ionizing flux is required to calculate the heating of the planet exosphere, and also more accurately determine the mass-loss rate, since the photo-ionisation rate and mass loss rate are highly correlated in transit observations \citep{Bourrier2013}. 

Accurate calculations of atmosphere heating and mass loss rates are required in order to constrain the energetic efficiency of exoplanet evaporation, which is important, as evaporation may have effects on the observed population of exoplanets \citep[e.g.][]{Davis2009, Jackson2012}. 

\begin{figure}
\begin{center}
\includegraphics[width=\columnwidth]{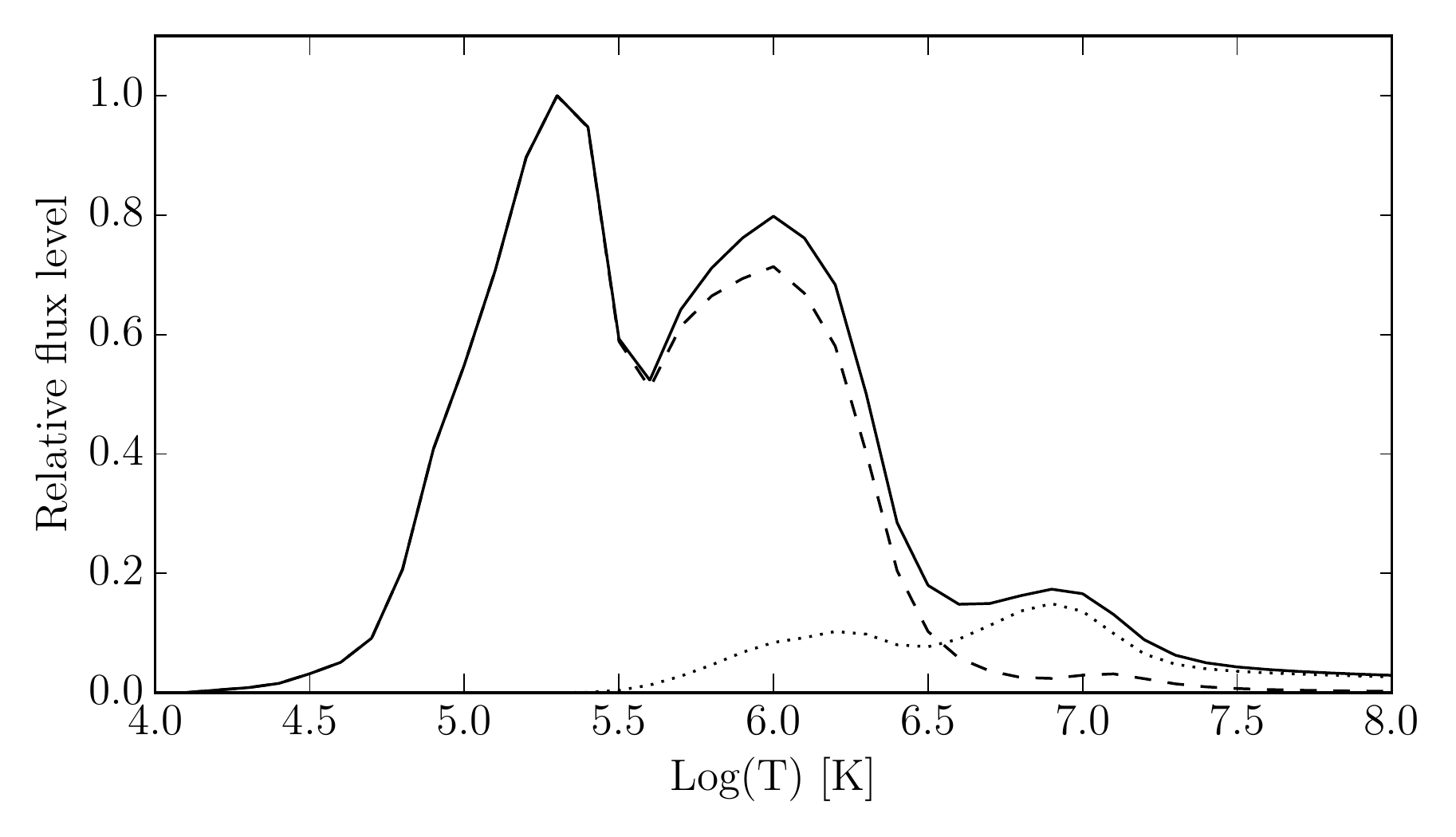}
\vspace{-20pt}
\caption{Relative emission of ionising radiation as a function of plasma temperature. Dotted line - X-ray flux (5--100 \AA), Dashed line - EUV flux (100--920 \AA), solid line - Total/XUV (5--920 \AA)}
\vspace{-20pt}
\label{fig:radloss}
\end{center}
\end{figure}

\section{Observations}

Ultraviolet data is obtained from \cite{France2010}, who use data from the \emph{HST} COS instrument (\emph{HST} GTO program 11534). Lines have been selected from their line list which are uncontaminated, have formation temperatures $> 10^{4.1}$ K, and which appear in the CHIANTI (v8.0) spectral database \citep{Dere1997,DelZanna2015}. The resulting line list is shown in table \ref{tab:linelist}.

HD\,209458 is listed as a detected source in the 3XMM DR5 catalogue (detid 104044501010110). The detection is based on observations taken in 2006 (Obs. ID 0404450101, PI Wheatley). The observations were made with the MEDIUM filters on all cameras, the PN was in FULL FRAME mode and the MOS cameras were in LARGE WINDOW mode to eliminate optical loading. The source detection is significant at over $3\sigma$ in the 0.2 - 0.5 keV band, but is not significantly detected at higher energies. This is not an unexpected characteristic for a stellar source, since the flux will be dominated by soft X-rays. As we report in section \ref{sec:results}, the corresponding flux is significantly higher than the upper limits reported by \cite{SanzForcada2011}, who analysed the same data. We are unsure of the cause of the discrepancy.

Previous claims of the detection of the flux of HD\,209458 \citep{Penz2008,Kashyap2008} are based on earlier \emph{XMM-Newton} data that were taken with the THIN optical blocking filters (Obs. ID 0130920101, PI Bertaux). For a bright target like HD\,209458b this can lead to contamination of the X-ray image by optical loading. We have analysed the spectrum of the earlier data and confirmed that the X-ray spectrum is unphysically soft, which is characteristic of optical contamination.

We use the reported count rates for the PN and MOS instruments in the five energy bands defined in the 3XMM catalogue.

\begin{center}
\begin{table}{
\caption{line strengths used in this work, obtained from \citet{France2010}, the max formation temperature is calculated using CHIANTI (v8.0)}
\begin{center}
\begin{tabular}{l c c c c c c}
\hline
\hline
Species & $\lambda_{\text{rest}}$ & Line Flux & T max\\
& (\AA) & ($10^{-16}$ erg cm$^{-2}$ s$^{-1}$) & \\
\hline
  \mbox{Si\,{\sc iii}} & 1206.50 & $12.89 \pm 0.35$ & 4.8\\
  \mbox{O\,{\sc v}} & 1218.34 & $\ 3.97 \pm 0.15$ & 5.4\\
  \mbox{N\,{\sc v}} & 1238.82 & $\ 1.08 \pm 0.58$ & 5.3\\
  \mbox{N\,{\sc v}} & 1242.80 & $\ 0.26 \pm 0.04$ & 5.3\\
  \mbox{N\,{\sc v}} & 1242.80 & $\ 1.30 \pm 1.07$ & 5.3\\
  \mbox{S\,{\sc ii}} & 1259.52 & $\ 1.68 \pm 0.54$ & 4.4\\
  \mbox{Si\,{\sc ii}} & 1264.74 & $\ 2.11 \pm 0.11$ & 4.1\\
  \mbox{Si\,{\sc iii}} & 1298.95 & $\ 0.63 \pm 0.09$ & 4.8\\
  \mbox{Si\,{\sc ii}} & 1309.28 & $\ 3.15 \pm 0.48$ & 4.1\\
  \mbox{C\,{\sc ii}} & 1334.53 & $\ 8.04 \pm 0.25$ & 4.3\\
  \mbox{C\,{\sc ii}} & 1335.71 & $16.06 \pm 0.14$ & 4.3\\
  \mbox{Si\,{\sc iv}} & 1393.76 & $\ 9.47 \pm 0.40$ & 4.9\\
  \mbox{Si\,{\sc iv}} & 1402.77 & $\ 5.11 \pm 0.49$ & 4.9\\
  \mbox{Si\,{\sc ii}} & 1526.71 & $\ \ \ 1.19 \pm 0.072$ & 4.1\\
  \mbox{Si\,{\sc ii}} & 1533.43 & $\ 2.31 \pm 0.14$ & 4.1\\
  \mbox{C\,{\sc iv}} & 1548.19 & $10.54 \pm 0.38$ & 5.1\\
  \mbox{C\,{\sc iv}} & 1550.77 & $\ 7.07 \pm 0.96$ & 5.1\\
  \mbox{Fe\,{\sc ii}} & 1559.08 & $\ 1.98 \pm 0.29$ & 4.1\\
  \mbox{Fe\,{\sc ii}} & 1563.79 & $\ 1.38 \pm 0.74$ & 4.1\\
  \mbox{Fe\,{\sc ii}} & 1570.24 & $\ 0.64 \pm 0.40$ & 4.1\\
\hline
\end{tabular}
\end{center}
\label{tab:linelist}
}
\end{table}
\end{center}

\section{Method}

\subsection{Differential Emission Measure}

The differential Emission Measure is a powerful diagnostic for an optically thin emitting plasma, such as a stellar corona. Essentially a measure of electron density along all lines of sight as a function of temperature, it can be used to characterize the atmosphere of a star. The DEM is defined as
\begin{equation}
\varphi(T) = {N_e}^2 \frac{dh}{dT}
\end{equation}
Where $N_e$ is the electron number density, T is the plasma temperature and h is the distance taken along the line of sight. The DEM is directly proportional to the strength of an emission line at a given temperature, so with a wide enough baseline of high temperature lines a near continuous DEM can be constructed. Combining a DEM with a suitable plasma model can recover the entire high energy spectrum, including continuum processes. In this work we use CHIANTI (v8.0) \citep{Dere1997,DelZanna2015}. HD\,209458 is a G class star with an [Fe/H] of $0.02 \pm 0.03$ \citep{Santos2004}, so we assume the corona has solar abundances \citep{Schmelz2012}.

\begin{figure*}
\begin{center}
\includegraphics[width=0.9\textwidth]{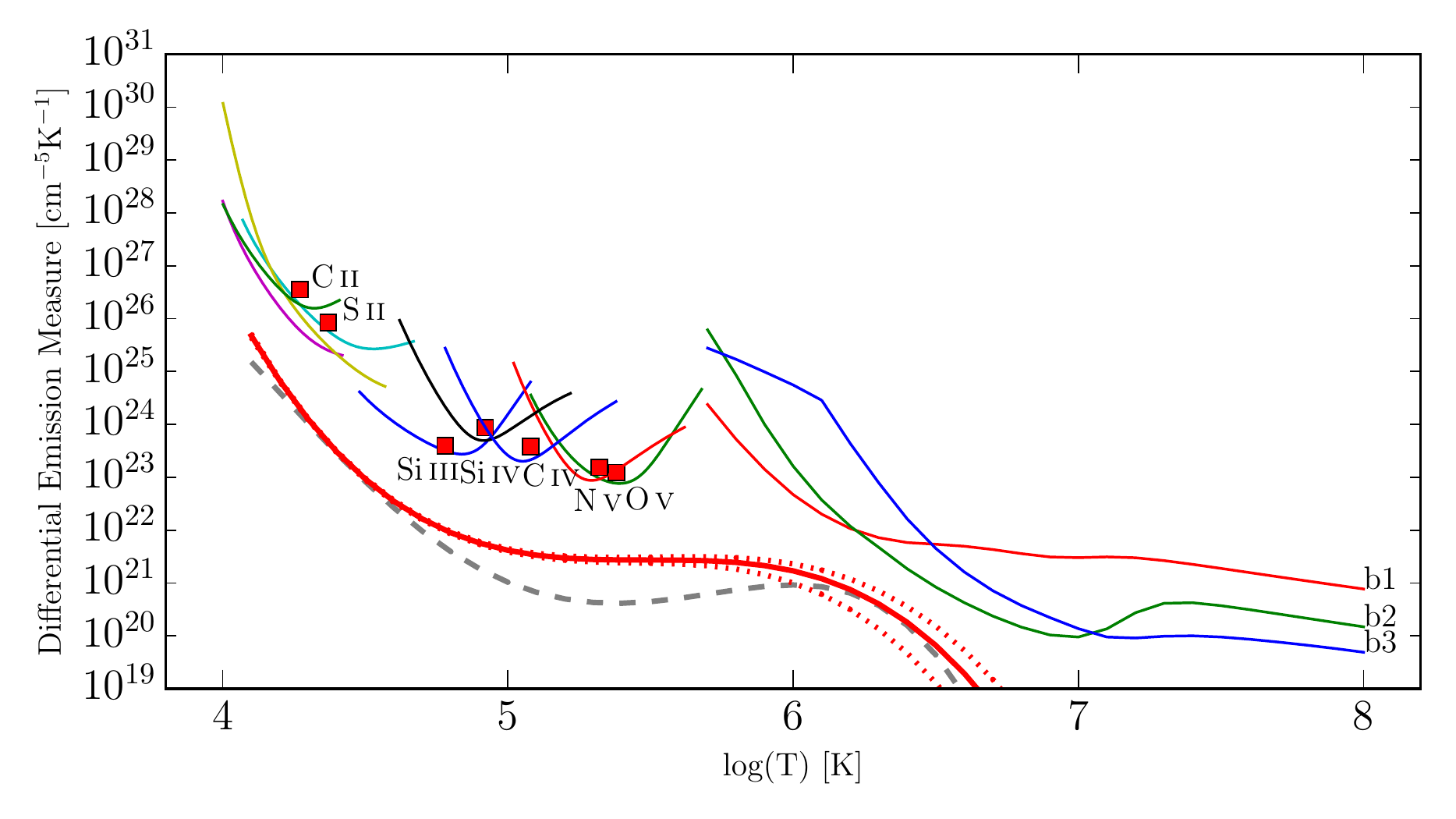}
\vspace{-10pt}
\caption{ The best-fitting DEM (red) with 3 $\sigma$ limits (red dotted). At the top left, the coloured lines are the emission loci, representing the constraints from the individual species contributing to the UV spectrum. The red squares are the average of the emission loci over a 0.3 dex temperature range. The emission loci and X-ray flux limits are the emission measure required for a single temperature model, and represent hard limits to the DEM. band 1 is 0.2--0.5 keV. Since no X-ray flux was detected in \emph{XMM-Newton} band 2 (0.5--1.0 keV) or band 3 (1.0--2.0 keV) the 3 $\sigma$ upper limit is plotted. The higher energy bands 4 and 5 do not impose any additional constraints, so are not plotted. A quiet solar DEM is plotted for comparison (grey dashed)\citep{Kretzschmar2004}}
\label{fig:DEM}
\end{center}
\end{figure*}

Figure \ref{fig:radloss} shows the relative strength of hydrogen ionising flux ($ \lambda <912$ \AA) emitted from the plasma as a function of temperature. It demonstrates the temperature regions that are most important to the level of ionizing radiation, and hence exospheric heating. The most potent temperatures are between $\approx10^{4.75}$ and $10^{6.5}$ K, with about half the integrated area $< ${\raise.17ex\hbox{$\scriptstyle\sim$}}$10^{5.7}$ K. This is the temperature region which can be well characterized and constrained by lines in the UV, such as those measured by \cite{France2010}, listed in Table \ref{tab:linelist}. Above $10^{5.7}$ K the strongest constraint is the soft X-ray flux. Combining these two constraints can allow the whole temperature range to be characterized.

Model spectra are folded through the instrument Response Matrix calculated using the \mbox{\sc science analysis system (sas)} \citep{Gabriel2004}. At the low fluxes under consideration pile-up is not an issue, so the relation between absolute flux level and count level should be linear. To simulate the detectable flux from the Earth we simulate interstellar medium absorption using the T{\"u}bingen-Boulder absorption model \citep{Wilms2000} generated in \mbox{\sc xspec} \citep{Arnaud1996}. We use a \mbox{H\,{\sc i}} column density $N_{H} = 2.3 \times 10^{18}$, which was calculated by fitting the Ly\,$\alpha$ profile by \citet{wood2005}.

\subsection{Fitting procedure}

We assume that the DEM of the star can be described by a smoothly varying function, which appears to be an adequate description for the Sun and other well studied stars. For our model we use a 4th order Chebyshev polynomial. The DEM is used to calculate the UV line strengths and scale the contribution from 40 pre-calculated model spectra, log evenly spaced in the temperature range $10^{4.1}$ - $10^8$ K. The model spectra and \emph{XMM-Newton} Response Matrixes are used to give a simulated \emph{XMM-Newton} response for each instrument and each temperature. The count rates in five energy bands are calculated: 0.2--0.5 keV, 0.5--1.0 keV, 1.0--2.0 keV, 2.0--4.5 keV and 4.5--12 keV.

The line strengths and \emph{XMM-Newton} count rates are compared to the measured values and errors to produce a residuals vector. In X-ray bands where no detection was made we use a value of 0 and the count errors from the background rate.

A Levenbert Marquardt process is initially used to solve the minimization problem. We use an MCMC to calculate the formal errors on the DEM and luminosity in all energy bands of interest. We use the emcee implementation in Python \citep{Foreman-Mackey2012}. We use a burn in of 10$^{7}$ steps to ensure the solution has converged and is well behaved, and use a further 10$^6$ steps for parameter estimation.

We use the stellar and planetary parameters reported by \citet{Southworth2010}. We choose not to use parameters derived from the interferometrically measured stellar radius from \citet{Boyajian2014}, as the authors caution that the angular size of HD\,209458 is close to the resolution limit of the instrument, so may be effected by systematics. We use the distance of 49.63 $\pm$ 1.97 pc from \citet{VanLeeuwen2007}, and include the distance errors in our reported luminosities.

\begin{center}
\begin{table}{
\caption{Numerical values for our recovered differential emission measure for HD\,209458b}
\begin{center}
\begin{tabular}{c c}
\hline
\hline
Log(T) & Log(DEM)\\
\hline
4.1 & $25.7$\\
4.2 & $24.8$\\
4.3 & $24.1$\\
4.4 & $23.5$\\
4.5 & $23.0$\\
4.6 & $22.5$\\
4.7 & $22.2$\\
4.8 & $22.0$\\
4.9 & $21.8$\\
5.0 & $21.6$\\
5.1 & $21.5$\\
5.2 & $21.5$\\
5.3 & $21.4$\\
5.4 & $21.4$\\
5.5 & $21.4$\\
5.6 & $21.4$\\
5.7 & $21.4$\\
5.8 & $21.4$\\
5.9 & $21.3$\\
6.0 & $21.2$\\
6.1 & $21.1$\\
6.2 & $20.9$\\
6.3 & $20.6$\\
6.4 & $20.3$\\
6.5 & $19.8$\\
6.6 & $19.3$\\
6.7 & $18.7$\\
6.8 & $17.9$\\
6.9 & $17.0$\\
7.0 & $16.0$\\
7.1 & $14.9$\\
7.2 & $13.6$\\
7.3 & $12.1$\\
7.4 & $10.5$\\
7.5 & $8.6$\\
7.6 & $6.6$\\
7.7 & $4.4$\\
7.8 & $2.0$\\
7.9 & $-0.6$\\
8.0 & $-3.5$\\
\hline
\end{tabular}
\end{center}
\label{tab:DEM}
}
\end{table}
\end{center}

\section{Results}
\label{sec:results}

\begin{figure*}
\begin{center}
\includegraphics[width=\textwidth]{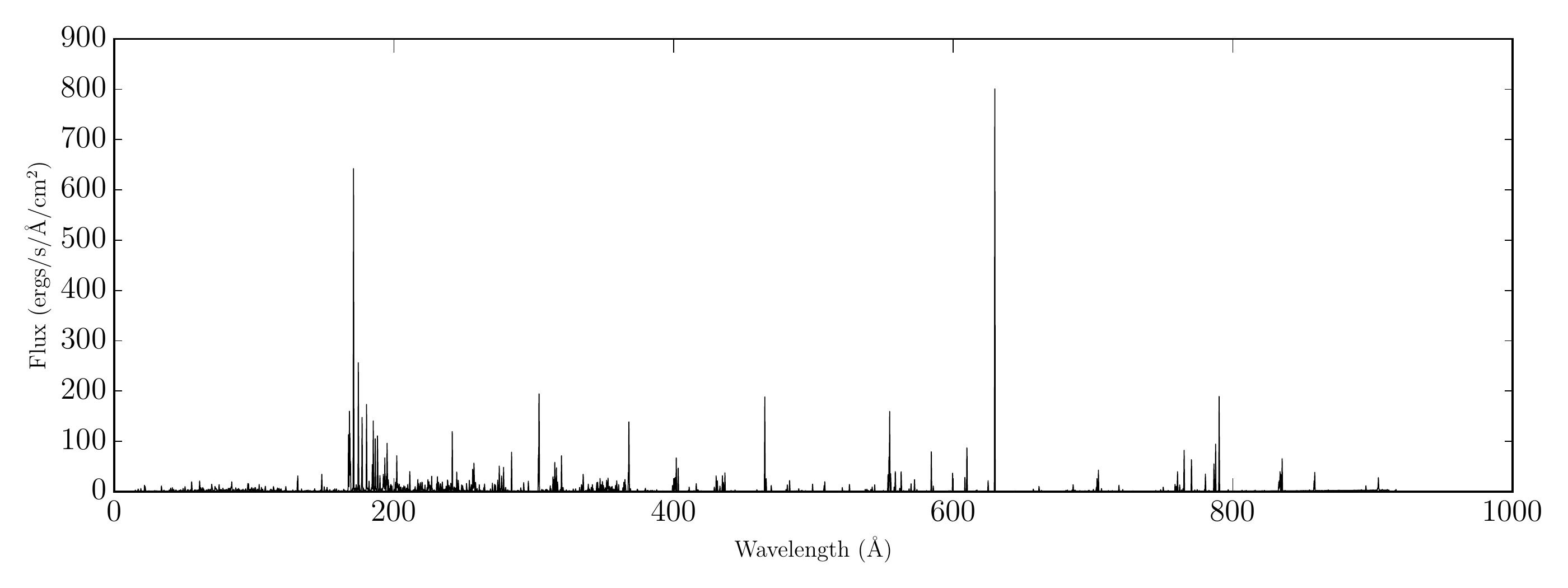}
\vspace{-20pt}
\caption{A recovered high energy spectrum for HD\,209458, scaled to the orbital distance of 0.047 AU.}
\label{fig:spectrum}
\end{center}
\end{figure*}

The best-fitting DEM is shown in Fig. \ref{fig:DEM}, with loci for each of the lines included in the fit, and the X-ray flux limit. The loci represent the value required to reproduce the line strength with a single temperature, so form an upper limit to the possible DEM. The spectrum in Fig. \ref{fig:spectrum} is generated from the best fit DEM. The XUV luminosity (defined as the integrated flux between 5 and 920 \AA) is 28.26$^{+0.05}_{-0.05}$ log(erg s$^{-1}$).

A 4th order polynomial was judged to provide the best fit to the data, higher orders became flexible enough to favour unphysical plasma distributions with multiple sharp peaks and drop-offs with temperature that did not resemble typical DEMs for the Sun or other solar type stars.

The electron density of the chromosphere/corona is uncertain; however, the major UV lines and broadband X-ray flux in this temperature range are only weakly dependent on plasma density. Some lines are susceptible to collisional de-excitation, but they are not important to the broadband fluxes. For our best-fitting model we assumed an electron density typical to the corona of the Sun, {\raise.17ex\hbox{$\scriptstyle\sim$}}$10^8$ cm$^{-3}$ \citep[e.g.][]{Fludra1999}. We find that varying the electron density over 5 orders of magnitude, between $10^{6}$ and $10^{11}$ cm$^{-3}$ has no significant effect on the total XUV flux of the model, all values agreeing to within 1$\sigma$ with the reported fluxes for $10^8$ cm$^{-3}$.

We also investigate the sensitivity of our results to errors in abundances. The metal abundances of the corona of the Sun are significantly different to the abundances of the photosphere. Different atmospheric regions of the Sun have enhanced abundances of some elements due to the First Ionisation Potential (FIP) effect \citep{Feldman2002}. In the Sun, atoms with FIP>11 eV have approximately the same abundance throughout the atmosphere, while low FIP elements with FIP<10 eV can be enhanced by factors of 3--5 in the hotter regions of the corona. HD\,209458 has a similar spectral type and activity level to the Sun, so would be expected to show a FIP effect of comparable strength \citep{Laming2004} but spatially resolved measurements of the atmospheric composition are unavailable. For our recovered spectrum we find that 43\% of the emitted XUV flux is from ions of iron alone, which has a measured photospheric abundance that is within 1 $\sigma$ of the solar value \citep{Santos2004}, giving confidence that the important coronal abundances are likely to be very similar to those of the Sun. The next highest contributor to XUV flux is oxygen, with 20\% of the total flux. Oxygen is a high FIP element and has little scatter in abundance for solar type stars with solar [Fe/H] \citep{Mena2015}. All other elements contribute significantly less than oxygen, so any errors in abundances either intrinsic or due to the FIP effect are unlikely to alter our total XUV flux by a significant amount.

\section{Discussion}

\renewcommand{\arraystretch}{1.5}
\begin{center}
\begin{table}{
\caption{The flux of our recovered EUV spectrum in several bands, compared to the Sun. $F_{XUV}$ is the flux received by HD\,209458b at 0.047 AU}
\begin{center}
\begin{tabular}{c c c c}
\hline
Spectrum & Log $L_X$ & Log $L_{EUV}$ & Log $F_{XUV}$\\
 & (erg s$^{-1}$) & (erg s$^{-1}$) & (erg s$^{-1}$ cm$^{-2}$)\\
\hline
Mean Sun & 27.38 & 27.93 & 3.28\\
Quiet Sun & 26.73 & 27.70 & 2.99\\
HD\,209458${^1}$ & <26.40 & <27.74 & <3.06\\
HD\,209458${^2}$ & 27.08 $^{+0.07}_{-0.07} $ & 28.23 $^{+0.05}_{-0.05} $ & 3.50 $^{+0.05}_{-0.05} $\\
\hline
\multicolumn{4}{l}{$^1$ \citet{SanzForcada2011}}.\\
\multicolumn{4}{l}{$^2$ This work.}\\
\end{tabular}
\end{center}
\label{tab:fluxes}
}
\end{table}
\end{center}
\renewcommand{\arraystretch}{1.0}

\subsection{Comparison to previous work}

The recovered DEM appears qualitatively similar to DEM measured for the quiet Sun, to demonstrate this we overplot the DEM found by \citep{Kretzschmar2004} in Figure \ref{fig:DEM}. The resemblance is quite striking, particularly in the low and high-temperature regions. In order to make a direct comparison of the fluxes, we generate a synthetic spectrum from this quiet Sun DEM using the same plasma model as for HD\,209458. Integrating the synthetic quiet Sun spectrum in the ROSAT bandpass (5--125\AA) gives a log flux of 26.8, which matches the solar minimum activity value of 26.8 reported for the ROSAT band reported by \citet{Judge2003}. We also compare our fluxes to those of the mean solar irradiance spectrum reported by \citet{Ribas2005}. We find that the X-ray flux falls between the quiet and mean Sun, but the total ionising flux (5--920\AA) flux of HD\,209458 is slightly higher than that of the mean Sun, by 0.2 dex.

\citet{SanzForcada2011} estimate the XUV flux of HD\,209458 by fitting an emission measure model to an X-ray upper limit measured from the same dataset, but with a broader energy band, and extrapolating to lower temperatures. We find that their results for total high energy flux are not consistent with ours. They estimate L$_{X}$ (5--100 \AA) and L$_{EUV}$ (100--920 \AA) upper limits of 26.40 and 27.74, which are significantly lower than the results from our recovered spectra at $27.08^{+0.07}_{-0.07}$ and $28.23^{+0.05}_{-0.05}$. Our results agree on the ratio of X-ray to EUV, which suggests our coronal models are compatible. The absolute {\raise.17ex\hbox{$\scriptstyle\sim$}}0.6 dex offset is due to their very low upper limit on the X-ray flux, which is significantly fainter than the flux reported in the 3XMM catalogue. Estimating the luminosity directly from the reported 3XMM flux gives a 0.2--12 keV luminosity of 26.8. This bandpass does not include all the soft X-ray flux of this very soft source, which is why our final recovered flux is slightly higher. The 3XMM source is at the expected position of HD\,209458b, and there are no other spurious sources detected nearby, so we believe this detection is genuine and that the X-ray flux is significantly higher than reported by \citet{SanzForcada2011}.

In our model 93\% of the XUV radiation is in the EUV band. \citet{Owen2012} compare the effects of X-ray and EUV radiation on the evaporation of exoplanets, and find that for planets like HD\,209458b, EUV radiation is dominant for atmosphere erosion, though the location and efficiency of absorption is different compared to the X-ray flux. \citet{Ehrenreich2011} calculate mass-loss efficiencies for HD\,209458b and HD\,189733b, however, they consider only X-ray flux in their calculations, which neglects the majority of the total ionising flux.

The STIS measurements of deep metal absorption presented in \citet{Vidal-Madjar2004} are unresolved, so there is no velocity information, but the thermal broadening of gas at 10,000 K would not be sufficient to absorb such a high fraction of the relatively broad stellar line profiles. \citet{Koskinen2013a} suggest that this can be explained by a metalicity enhancement of a factor of 3--5 in the atmosphere of HD\,209458b, accompanied by the XUV flux being greater than a factor of 2 higher than the mean solar flux, which is not consistent with our recovered XUV flux. This discrepency may be a sign that superthermal processes are at work in the atmosphere of HD\,209458b \citep{Ballester2015}.

\citet{Guo2015} study the effects of spectral energy distribution of EUV radiation on the structure of hot Jupiter atmospheres and their evaporation. They define an EUV spectral index $\beta$ which is the ratio of the flux in a 50--400\,\AA$ $ band to the flux in a 50--900\,\AA$ $ band. The value of $\beta$ for our recovered stellar spectrum is $0.64^{+0.02}_{-0.02}$. Our synthetic quiet Sun spectrum gives a value of beta of 0.74, which matches observed spectra of the Sun during periods of low activity. As a G type star, it is perhaps not surprising that our recovered spectrum for HD\,209458 has a qualitatively similar spectral shape to that of the quiet Sun. \citet{Guo2015} show that the location of the H/H$^+$ transition in the atmosphere outflow of HD\,209458b depends strongly on the spectral index $\beta$. The location of this transition region itself is important to calculating the structure of the wind. Our measurement of this value is an important input to any future models of the evaporation and structure of HD\,209458b's atmosphere, both for driving escape and constraining the photo-ionisation of the wind.

\subsection{Energy-limited mass-loss}

Energy-limited mass-loss can be used to calculate an upper limit to the evaporation rate of an exoplanet. The only assumption is that some fraction of incoming high-energy flux, $\eta$, is converted into work done against the gravitational potential of the planet \citep{Watson1981}. Using the formulation from \cite{Erkaev2007}, the equation for energy-limited mass-loss is:

\begin{align}
\dot{M} &= \frac{\eta \pi F_{XUV} \alpha^2 R_{P}^3}{G M_p K} \\
K &= \left(1 - \frac{3}{2 \xi} + \frac{1}{2 \xi ^3} \right) \\
\xi &= \frac{R_{Hill}}{R_{p}}
\end{align}

where $F_{XUV}$ is the X-ray and EUV flux incident on the planet, G is the gravitational constant and $M_p$ is the mass of the planet. The formula includes a K factor to account for the reduced energy required to reach the Hill sphere, $R_{Hill}$ where the gas is effectively unbound from the planet. For a close-in hot Jupiter like HD\,209458b, this factor can increase the predicted mass-loss rate by upwards of 50\%. The $\alpha$ factor accounts for the increased cross section of the planet to XUV radiation, since it is typically absorbed very high in the atmosphere. \newpage
This value can be derived using a simple analytical structure model of the atmosphere  \citep{Murray-Clay2009}. The optical depth of 1 for incoming EUV radiation will occur at approximately 1 nanobar, which is 20 scale heights above the 1 bar optical radius of the planet. Assuming the dayside atmosphere is approximately isothermal puts the XUV absorption radius at {\raise.17ex\hbox{$\scriptstyle\sim$}}1.1R$_p$, making $\alpha$ 1.1. X-rays are absorbed deeper in the atmosphere, but they make up a comparatively small contribution to the flux.

The mass-loss efficiency, $\eta$, is unknown, which is a major limitation for models of mass loss. There are calculations of the expected value of $\eta$, but they are far from precise. Based on modelling of Hot Neptunes, \cite{Owen2012} suggest that $\eta$ should be between 0.05 and 0.2. At very high fluxes, the energy transfer transitions to a recombination limited regime where mass-loss scales with $F_{XUV}^{1/2}$, and the efficiency decreases, since a greater proportion of energy is being lost to recombination \cite{Murray-Clay2009}. \citet{Shematovich2014} simulated absorption of XUV radiation in the lower thermosphere of HD\,209458b and find $0.10 < \eta < 0.15$. \citet{Salz2016} perform hydrodynamic simulations of mass-loss and XUV absorption up to the Roche-lobe height, and find $\eta = 0.21$ for the case of HD\,209458b, though the exact efficiency depends on the input spectrum. \citet{Koskinen2013} find higher heating efficiencies of 0.4--0.6 for XUV fluxes {\raise.17ex\hbox{$\scriptstyle\sim$}}450 times today's solar flux, however, for planets with lower incident XUV flux, IR cooling from photochemically produced H$_{3}^{+}$ becomes important \citep{Koskinen2007}, and can decrease heating efficiency to closer to 0.15. An efficiency of 0.15 has also been found to be appropriate for the case of Venus \citep{Chassefiere1996}. The planets magnetic field could also effect the efficiency, and will require magnetohydrodynamic simulations to model \citep[e.g.][]{Owen2014}. A recovered high energy spectrum will allow better estimates both of the input energy and mass loss rates of the evaporating atmosphere, improving the observational constraints on $\eta$.

Assuming $\eta=0.2$ our best-fitting DEM incurs a mass-loss rate of $3.8 \pm 0.2 \times 10^{10}$ g s$^{-1}$, compatible with the reported values of {\raise.17ex\hbox{$\scriptstyle\sim$}}$10^{10}$ g s$^{-1}$ in \citet{Vidal-Madjar2003,Vidal-Madjar2004} and the models of \cite{Owen2012}, but significantly lower than the 8--40 $\times 10^{10}$ g s$^{-1}$value reported by \cite{Linsky2010}. The efficiency would have to be at least 0.42 to reach the lower end of this range, and the higher value cannot be reached without exceeding 100\% efficiency, which is clearly unphysical. This implies that the mass-loss rates have perhaps been systematically overestimated, as \citet{Ballester2015} suggest. We emphasise that a strength of our analysis is that the ultraviolet line strengths we use in our analysis are taken from the same dataset that is used to calculate the planet mass-loss rates, so stellar variability is not a concern.

In reality, energy-limited mass-loss ignores a number of important physical effects, and should be regarded as an upper limit. 3d models are required to properly model higher order features such as the shape of the outflow and self-shielding effects, but require an accurate XUV spectrum as a model input. \citet{Bourrier2013} show that the estimated hydrogen escape rate and the level of ionising flux are highly related. Since only neutral hydrogen is visible in the Ly\,$\alpha$ absorption, it is important to understand the ionising flux in order to correctly infer the true mass-loss rate. The ionizing flux we measure for HD\,209458 is {\raise.17ex\hbox{$\scriptstyle\sim$}}$3\times$ higher than the solar minimum value, so in their model is most consistent with a mass-loss rate of 10$^{9}$ g s$^{-1}$, and is inconsistent at the $2 \sigma$ level with higher mass-loss rates.

\section{Conclusions}
We have constructed a synthetic coronal spectrum for HD\,209458 using ultraviolet line strengths and X-ray fluxes. We find that the star is comparable in activity levels to the Sun, with a total XUV flux of 10$^{28.26}$ ergs s$^{-1}$. Assuming a mass loss efficiency of 0.2, this could cause the atmosphere of HD\,209458b to evaporate at a rate of $3.8 \pm 0.2 \times 10^{10}$ g s$^{-1}$, which is significantly lower than the estimated value of 8--40 $\times 10^{10}$ g s$^{-1}$ of \cite{Linsky2010}. The higher end of this range cannot be reached without exceeding 100\% efficiency, which is not physical, and the efficiency would have to be at least 0.42 to reach the low end. Models of atmospheric efficiency typically do not find efficiencies greater than 0.2 \cite[e.g.][]{Owen2012}, and it is often lower. This suggests that evaporation rates may have been overestimated, due to not accounting for stellar variability \citet{Ballester2015}, and likely compounded by degeneracies between outflow structure/ionization and total mass \citet{Bourrier2013}. Modelling the outflow correctly requires an input spectrum, as the neutral hydrogen that is observed through Ly\,$\alpha$ absorption is sculpted by radiation pressure and photoionisation rates, which are dependent on both the total XUV flux, but also its spectral shape \citet{Guo2015}. Our DEM recovery provides these important values for future models of the evaporation of HD\,209458b.

\section*{Acknowledgements}

We thank to Dr. John Pye for very useful comments. KB is grateful the Paul Scherrer Institute and ETH Zurich, Switzerland, where some of the early work for this paper was done. T.L. is supported by a STFC studentship. P.W. is supported by a STFC consolidated grant (ST/L000733/1).




\bibliographystyle{mnras}
\bibliography{bibliography}








\bsp	
\label{lastpage}
\end{document}